\documentclass[aps,prb,amsmath,amssymb,reprint]{revtex4-1}
\usepackage{graphicx}
\usepackage{dcolumn}
\usepackage{bm}
\newcommand{\degree}{\ensuremath{^\circ}}
\newcommand{\figsize}{7.8cm}

\usepackage{color}
\usepackage{xcolor}

\begin{document}
\title{Swift heavy ion irradiation of GaSb: from ion tracks to nano-porous networks} 
\author{C. Notthoff}
\email[]{christian.notthoff@anu.edu.au}
\affiliation{Department of Electronic Materials Engineering, Research School of Physics, The Australian National University, Canberra 2601 ACT, Australia.}

\author{S. Jordan}
\affiliation{Department of Electronic Materials Engineering, Research School of Physics, The Australian National University, Canberra 2601 ACT, Australia.}

\author{A. Hadley}
\affiliation{Department of Electronic Materials Engineering, Research School of Physics, The Australian National University, Canberra 2601 ACT, Australia.}

\author{P. Mota-Santiago}
\affiliation{Department of Electronic Materials Engineering, Research School of Physics, The Australian National University, Canberra 2601 ACT, Australia.}

\author{R. G. Elliman}
\affiliation{Department of Electronic Materials Engineering, Research School of Physics, The Australian National University, Canberra 2601 ACT, Australia.}

\author{W. Lei}
\affiliation{Department of Electronic Materials Engineering, Research School of Physics, The Australian National University, Canberra, Australia.}
\affiliation{School of Electrical, Electronic and Computer Engineering, The University of Western Australia, Crawley, Western Australia 6009, Australia.}

\author{N. Kirby}
\affiliation{Australian Synchrotron part of ANSTO, Melbourne, Australia.}

\author{P. Kluth}
\affiliation{Department of Electronic Materials Engineering, Research School of Physics, The Australian National University, Canberra 2601 ACT, Australia. }

\date{10/04/2020}

\begin{abstract}
Ion track formation, amorphisation, and the formation of porosity in crystalline GaSb induced by 185 MeV $^{197}$Au swift heavy ion irradiation is investigated as a function of fluence and irradiation angle relative to the surface normal. Rutherford backscattering spectrometry in channeling configuration and small-angle X-ray scattering reveal an ion track radius between 3 and 5 nm. The observed pore morphology and saturation swelling of GaSb films shows a strong irradiation angle dependence. Raman spectroscopy and scanning electron microscopy show that the ion tracks act as a source of strain in the material, leading to macroscopic plastic flow at high fluences and off normal irradiation. The results are consistent with the ion hammering model for glasses. Furthermore, wide-angle X-ray scattering reveals the formation of nano-crystallites inside otherwise amorphous GaSb after the onset of porosity.
\end{abstract}

\maketitle 
\section{Introduction}
GaSb is a narrow band-gap semiconductor with many technological applications, including laser diodes, high-frequency electronic devices \cite{JAP81}, high-efficiency infra red photodetectors, thermoelectric devices, thermo photovoltaics, and tandem concentrator solar cells \cite{PK2,PK3}. Porous semiconductors differ significantly in their physical and chemical properties from their bulk counterparts due to their microstructure that is often characterised by a large surface-to-volume ratio and small feature sizes. Exploring such properties, nano-fporous semiconductors have been identified as ideal building blocks for many optoelectronic, thermoelectric, thermo photovoltaic and sensor devices, and membranes for biological and chemical applications \cite{PK4, PK5, N2}. The controlled fabrication of porous semiconductors thus paves the way for the development of new materials with application-specific properties. Commonly used methods to prepare nano-porous semiconductors are sintering of nanoparticles \cite{PRL107}, electrochemical etching \cite{Feng1994,Foell2003}, and ion irradiation \cite{JPD42,PK6,PK7,PK9,PK10,PRB84}. However, all methods have their own challenges to control and tune the porosification process.
Electrochemical etching, for example, is typically only capable of rendering a thin layer of a few nanometers into a porous structure, with the exception of porous silicon\cite{porous_silicon,Beale1985}. It has been demonstrated previously that ion irradiation at low energies can lead to the formation of nano-porous structures in semiconductors such as GaSb, InSb, and Ge \cite{PK6,PK7,PK9,PK10,PRB84,JAP115}. The formation of porosity is attributed to clustering of vacancies that are generated during the elastic collisions when the material is irradiated by low energetic ions.\\
\indent
We have recently discovered the evolution of nano-porous structures in GaSb following swift heavy ion irradiation, where nuclear collisions become negligible and the interaction is dominated by electronic energy loss \cite{Kluth2014,REI2017}. The porous structures generated by swift heavy ion irradiation in GaSb are fundamentally different from those resulting from low-energy irradiation or electrochemical etching. The process allows fabrication of significantly thicker layers up to several micrometers, in contrast to a few nanometers typical for etching and low-energy irradiation. The swift heavy ion process is more efficient and furthermore enables the controlled fabrication of a new class of porous materials. There is only one publication \cite{PRB65} on ion track formation in GaSb under swift heavy ion irradiation and our own work on the mechanisms of porous structure formation \cite{Kluth2014,REI2017} at high fluences.\\
\indent
In this work, we present results on ion track formation and porosification in crystalline GaSb under 185 MeV Au swift heavy ion irradiation. The formation of amorphous tracks in an otherwise crystalline GaSb was investigated at low fluence ($\Phi \ll 1\times10^{13}$ ions/cm$^2$) using Rutherford backscattering spectrometry in channeling geometry, small-angle X-ray scattering, and Raman spectroscopy. Furthermore, the effect of ion irradiation at high fluences ($\Phi > 1\times10^{13}$ ions/cm$^2$) and non perpendicular ion incidence with respect to the surface is investigated using high-resolution scanning electron microscopy and grazing incidence wide-angle X-ray scattering.
\section{Experimental\vspace*{-0.3cm}}
Single-crystal GaSb layers grown on (001) InP substrates by metal organic chemical vapor deposition (MOCVD) and bulk GaSb single-crystal wafers were irradiated at room temperature with 185 MeV $^{197}$Au ions at the Australian National University Heavy Ion Accelerator Facility. Irradiation was performed to fluences ranging from $5.6\times10^{11}$ to $2\times10^{14}$~ions/cm$^2$ and at angles of incidence relative to the surface normal between 0$\degree$~and 60$\degree$. The surface electronic energy loss and mean ion range estimated using SRIM-2013 \cite{srim2013} are 22.3 keV/nm and 16.5 $\mu$m, respectively. A layer thickness of approximately 2.4 $\mu$m was chosen for the MOCVD samples to ensure that the energy loss in the GaSb layers is dominated by electronic stopping \cite{Kluth2014}, even for irradiation angles up to 60$\degree$. Samples were studied using scanning electron microscopy (SEM), Raman spectroscopy, Rutherford backscattering spectrometry in channeling configuration (RBS/C), and small- and wide-angle X-ray scattering (SAXS/WAXS). For SEM analysis, the samples were cleaved and imaged in cross section to investigate the morphological changes. Samples irradiated under an angle relative to the surface normal were cleaved in the plane of irradiation and perpendicular to it (not shown).\\
\indent
Track formation and damage buildup at low fluences ($\Phi<1\times10^{13}$ ions/cm$^2$) was investigated with RBS/C using (001)-oriented bulk GaSb samples, 2 MeV He$^{2+}$ ions, and a surface barrier detector positioned at a scattering angle of 168$\degree$.\\
\indent
For the SAXS measurements at the Australian Synchrotron, thin-film samples irradiated to a fluence of $\Phi=3\times 10^{12}$~ions/cm$^2$ were used where the substrate was removed post irradiation by selective etching with HCl. The top surface of the GaSb film was protected by Kapton tape during 30 min etching, which was also used as support during the SAXS measurements. The SAXS measurements were performed in transmission geometry with an X-ray energy of 12 keV (wavelength $\lambda$ = 1.0332 \AA) and a sample-to-detector distance of 968 mm. Silver behenate and glassy carbon reference samples were used for $q$-space calibration and normalisation of the absolute scattering intensity, respectively. SAXS data were taken at room temperature with the ion tracks tilted by about 10$\degree$ with respect to the X-ray beam, using a Pilatus 1M detector.\\
\indent	
Thin-film samples irradiated at normal and $30\degree$ incidence with fluences between $5.6\times10^{12}$ and $8.8\times10^{13}$~ions/cm$^2$ were investigated with grazing incidence WAXS using the GaSb films on InP without any further preparation. The WAXS measurements were performed at $\alpha_i=1\degree$ incidence angle, an X-ray energy of 14 keV ($\lambda = 0.885601$ \AA), and a sample-to-detector distance of 476 mm using a Pilatus 200k Detector. A LaB$_6$ reference sample was used for $2\theta$-space calibration and to determine the instrumental broadening.
\section{Results\vspace*{-0.3cm}}
\subsection{Track formation and damage cross-section\vspace*{-0.3cm}}
\begin{figure}
 \includegraphics[width=\figsize]{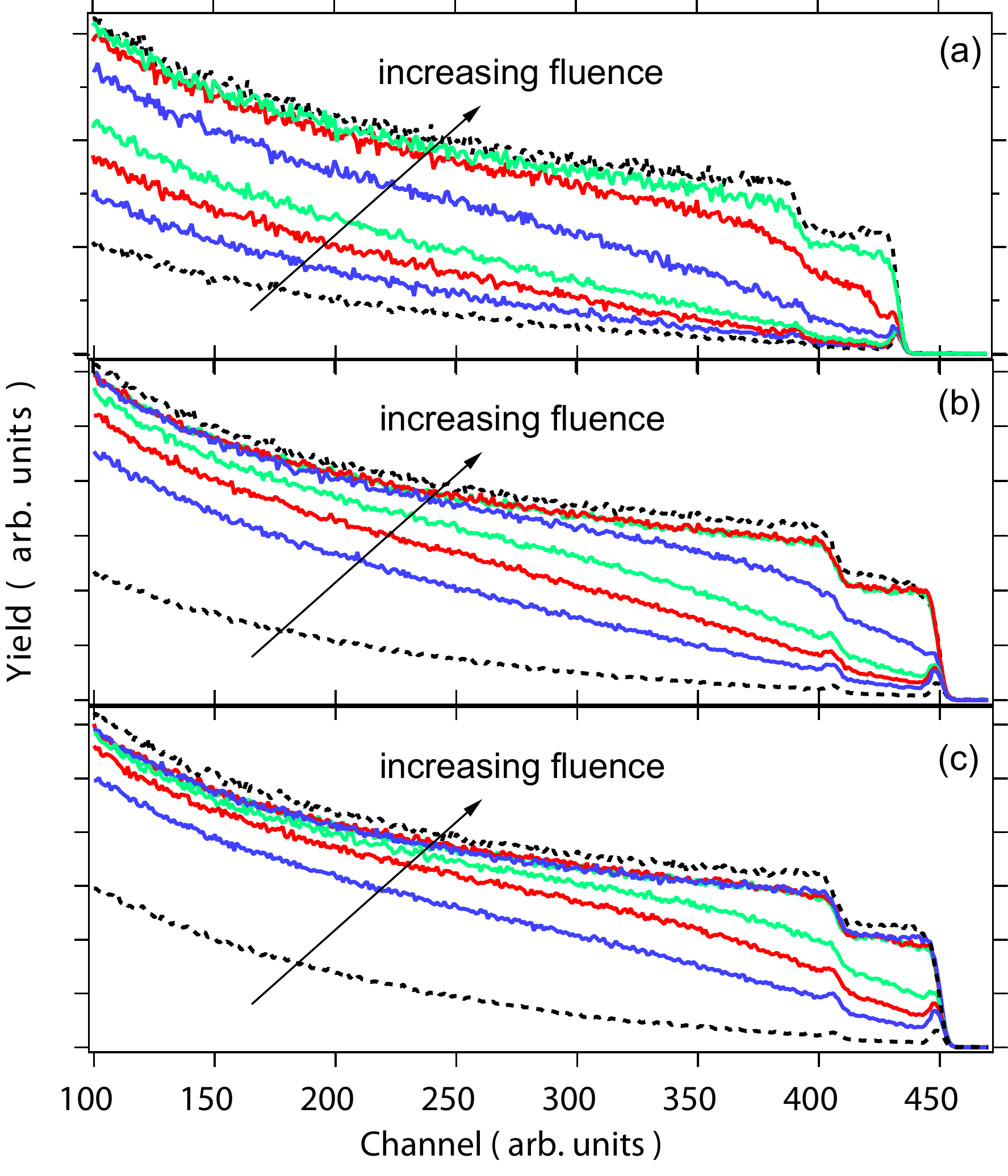}%
 \caption{RBS/C backscattering yield for samples irradiated with with 185 MeV Au ions at varying fluences (5.6$\times10^{11}$, 8.8$\times10^{11}$, 1.2$\times10^{12}$, 2.4$\times10^{12}$, 5.6$\times10^{12}$, and 8.8$\times10^{12}$ ions/cm$^2$) (a) normal to the surface, and tilted by (b) 30 $\degree$, and  (c) 60 $\degree$, relative to the surface normal. The dashed lines are channeled and randomly oriented reference measurements of a pristine GaSb sample.\vspace*{-0.5cm} \label{fig:rbs_raw_0_30_60}}%
\end{figure}
Bulk GaSb samples irradiated with fluences ranging from $5.6\times10^{11}$ to $8.8\times10^{12}$ ions/cm$^2$ were investigated by RBS/C. The analysis depth for 2 MeV He$^{2+}$ ions in GaSb is $\sim$ 2 $\mu$m, almost 10 times smaller than the range of 185 MeV $^{197}$Au ions used for irradiation. Therefore, the samples are expected to appear uniform over the RBS/C measurement depth. Figure \ref{fig:rbs_raw_0_30_60} shows RBS/C data for samples irradiated at normal (a), 30\degree~(b), and 60\degree~(c) incidence angle with $5.6\times10^{11}$, $8.8\times10^{11}$, $1.2\times10^{12}$, $2.4\times10^{12}$, $5.6\times10^{12}$, and $8.8\times10^{12}$ ions/cm$^2$. A pristine, (001)-oriented, single-crystal GaSb sample was measured in channeling and random configuration for reference (shown as dashed lines in Fig. \ref{fig:rbs_raw_0_30_60}). With increasing swift heavy ion irradiation fluence, we observe a continuous change of the backscattering yield from crystalline to random, which is a direct measure of the increase in disorder/damage in the material. For quantitative\\ analysis of the damage as a function of fluence $\Phi$, we have calculated the volume fraction of damaged material by 
\begin{eqnarray}
\small
f_d(\Phi)=\frac{Y(\Phi)-Y_{c}}{Y_{r}-Y_{c}}~,
\label{eq:RBS_damage}
\end{eqnarray}
with the backscattering yields integrated between channels 100 and 470 for the different irradiated samples $Y(\Phi)$, the reference sample in channeling configuration $Y_c$, and the random measurement $Y_r$.
 \begin{figure}
 \includegraphics[width=\figsize]{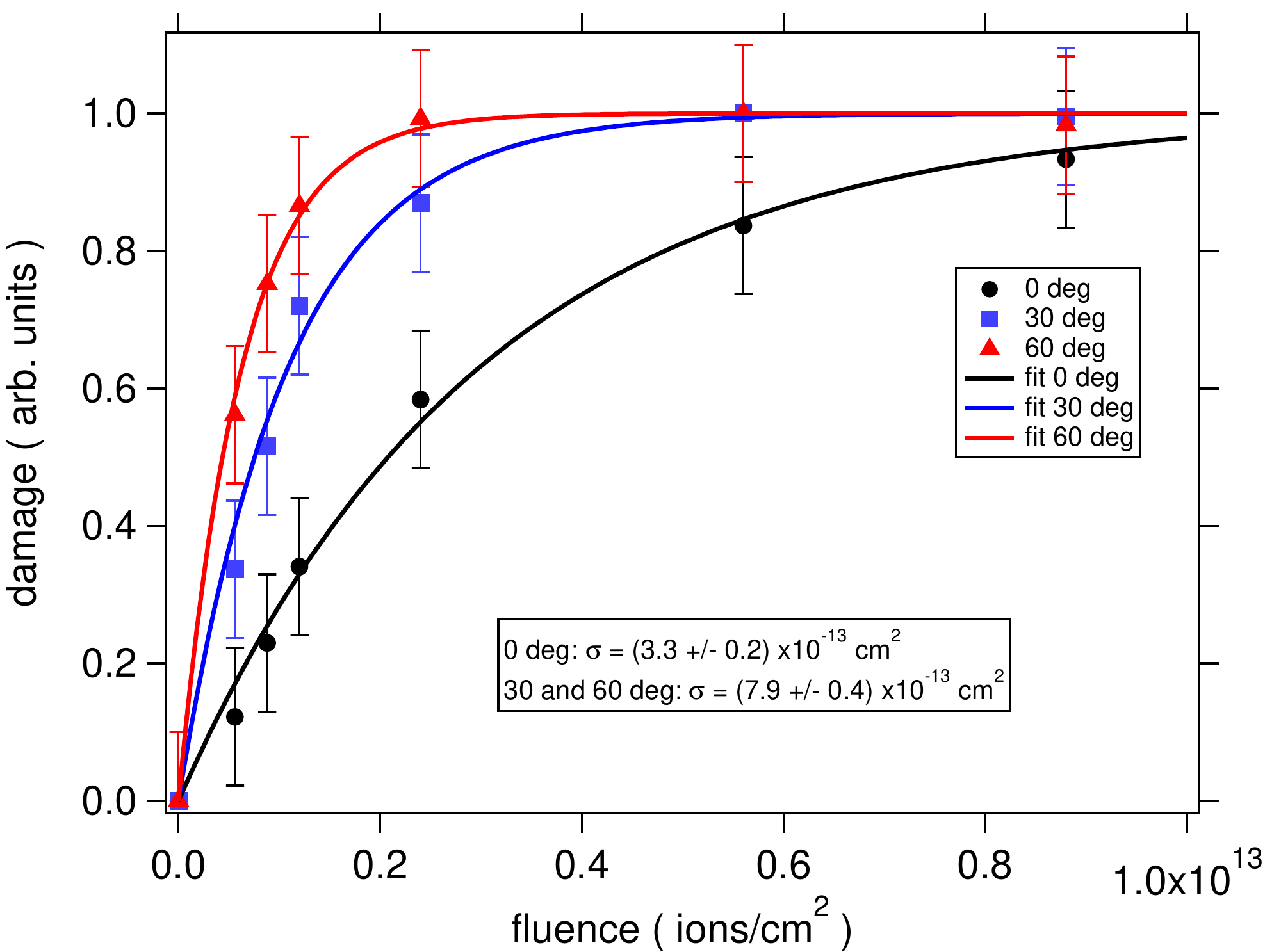}%
 \caption{Damage buildup as a function of fluence for samples irradiated normal to the surface (black dots), 30 $\degree$ (blue squares), and  60 $\degree$~(red triangles) relative to the surface normal. Solid lines are fits of Eq. (\ref{eq:RBS}) to the data.\vspace*{-0.5cm}\label{fig:rbs_eval_0_30_60}}%
 \end{figure}
Figure \ref{fig:rbs_eval_0_30_60} shows the resulting volume fractions of damaged material as a function of fluence determined from the RBS/C spectra shown in Fig.~\ref{fig:rbs_raw_0_30_60}.
The volume fraction of damaged material as a function of fluence $\Phi$ follows a Poisson law \cite{IEEE60}:
\begin{eqnarray}
\small
f_d(\Phi)=1-e^{-\sigma \frac{\Phi}{ \cos(\theta)}}~,
\label{eq:RBS}
\end{eqnarray}
with the damage cross section $\sigma$, and the swift heavy ion incidence angle $\theta$.
The $\cos(\theta)$ term in Eq. (\ref{eq:RBS}) accounts for the extended path of the ions when irradiation is performed under an angle $\theta$ relative to the surface normal.
The damage buildup for irradiation at 30$\degree$ and 60$\degree$ can be fitted well with a single cross section of $\sigma=(7.9\pm0.4)\times10^{-13}$~cm$^2$. For irradiation at normal incidence, however, we observe a significantly reduced cross section of $\sigma=(3.3\pm0.2)\times10^{-13}$~cm$^2$.
Assuming a cylindrical interaction volume without any inner structure, the cross section $\sigma$ can be converted to an equivalent ion track radius of $R_{RBS}\approx3$~nm and 5~nm for normal and off-normal incidence, respectively. We attribute the smaller damage cross section to a reduction of energy loss $S_e$ due to the channeling effect present at normal incidence \cite{channel1,channel2}. These results differ significantly from those by Szenes \textit{et al.} \cite{PRB65}, who report, based on transmission electron microscopy (TEM) investigations, no track formation for Pb irradiation at 0.85 MeV/u corresponding to an energy loss of 21.9 keV/nm, which is similar to the 22.3 keV/nm for 0.94 MeV/u Au ions used here. Furthermore, the track radii are significantly larger than the 1.8 nm reported for Pb irradiation at 1.85 MeV/u in the same study, based on RBS.
\subsection{Synchrotron based small angle X-ray scattering\vspace*{-0.4cm}}
\begin {figure}
\includegraphics [width=\figsize ]{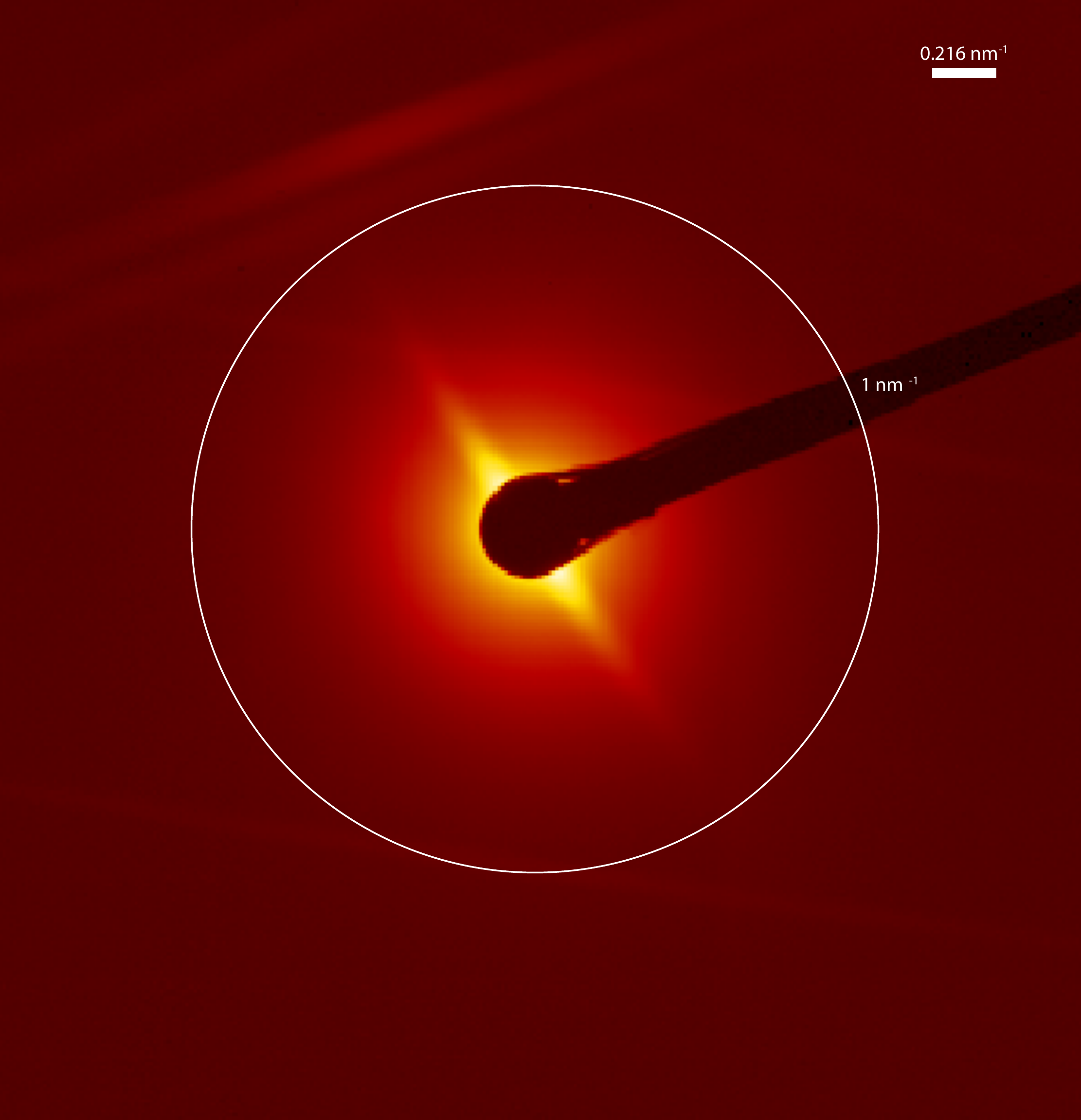}\\%
\caption {SAXS image of a GaSb sample irradiated with 3$\times 10^{12}$ Au ions/cm$^{2}$ at 185 MeV and normal incidence, after selective etch removal of the InP substrate. The white ring indicates the radial $q$ value, $q=1$ nm$^{-1}$. \vspace*{-0.5cm}\label{fig:SAXS_raw}}%
\end {figure}
Figure \ref{fig:SAXS_raw} shows a transmission SAXS image of a freestanding GaSb film irradiated with $3\times10^{12}$ ions/cm$^2$, where the InP substrate was selectively removed by a post irradiation HCL etching step. At a fluence of $3\times10^{12}$ ions/cm$^2$ the overlap of the ion tracks is small enough to interpret the tracks as separate, well-aligned, cylindrical, amorphous inclusions in an otherwise crystalline matrix. As discussed later, no macroscopic porosification is observed at fluences below $5\times10^{12}$ ions/cm$^2$, which is consistent with no observation of scattering from nano-sized pores in the SAXS experiment. SAXS has been extensively used previously to study ion tracks in various materials \cite{PRB83a,Rodriguez2012}. The sample and therefore the long axis of the ion tracks was tilted by $\approx 10\degree$ relative to the X-ray beam, resulting in clear, well-developed streaks typical for ion tracks with high aspect ratios \cite{PRB83a,Rodriguez2012}.
\begin{figure}[h]
	\includegraphics[width=\figsize]{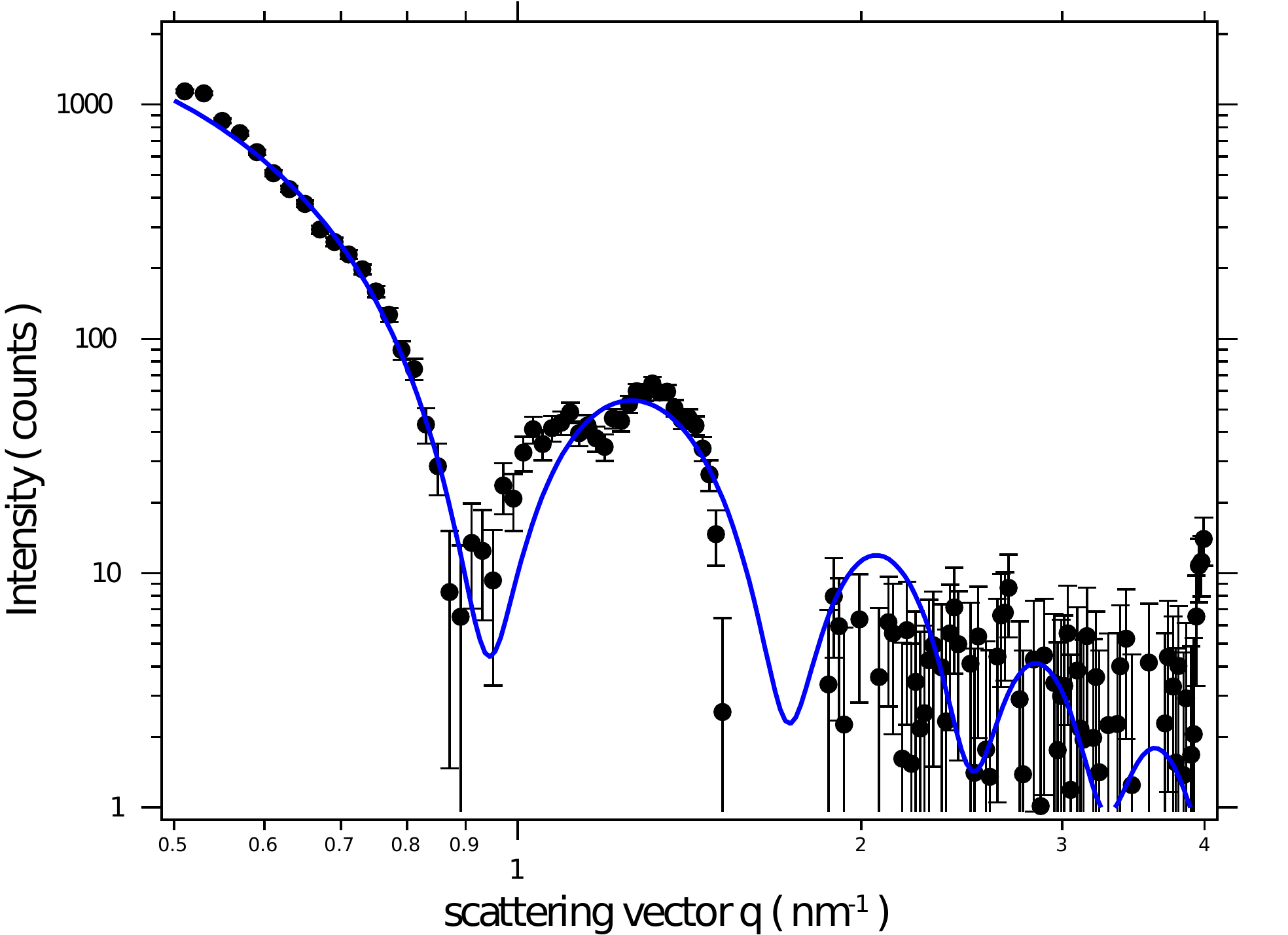}%
	\caption{Background-corrected SAXS data extracted from Fig. \ref{fig:SAXS_raw} (black dots) and a simple oriented cylinder model fit (blue line).\vspace*{-0.6cm}\label{fig:SAXS_fit}}%
\end{figure}
For quantitative analysis of the SAXS data, the scattering intensity along the streak is extracted and background corrected by subtracting the isotropic scattering contribution from the matrix. The anisotropic scattering of the ion tracks is extracted by applying a narrow mask along the streak, excluding all but the intensity of the streak, and azimuthal averaging of the 2D data in the $q$ range from $q_{min}=0.5$ nm$^{-1}$ to $q_{max}=4.0$ nm$^{-1}$. Using the same mask but rotated by about 5$\degree$~relative to the streak that excludes the anisotropic scattering of the ion tracks allows for the extraction of the background from the same data set \cite{Rodriguez2012}. The data reduction is performed with a custom developed python code \cite{mycode} which uses pyFAI \cite{pyfai} for data binning. Figure \ref{fig:SAXS_fit} shows the scattering intensities along the streak after background removal (black dots) together with a simple oriented cylinder model fit\cite{PRB83a,Rodriguez2012} (solid blue line):
\begin{eqnarray}
\small
I(q,R_{SAXS},\chi)=&\frac{1}{\eta}\frac{(2\pi \Delta \rho L)^2 N} {\sqrt{2\pi}\chi} \quad\quad\quad\quad\quad\quad\quad\quad\quad \nonumber \\ \times&  \int_{-\infty}^{\infty}dr \left| \frac{r~J_1(q~r)}{q}\right|^2e^{-\frac{(r-R_{SAXS})^2}{2\chi^2}}~,\quad
\end{eqnarray}
with the length of the ion tracks (sample thickness) $L=2.4$~$\mu$m, number of ion tracks $N=3\times10^{12}$, scattering length density difference $\Delta \rho=\rho_{track}-\rho_e$ between the ion track and bulk, and a conversion factor $\eta=3.26\times10^{-5}$ cm/counts from absolute intensities to detector counts obtained from a reference measurement of glassy carbon under the same experimental conditions \cite{absSAXS}.
The cylinder model agrees well with the experimental scattering intensities (except for a slight deviation at about 1 nm$^{-1}$ where part of the signal stems from a Kossel line). The fit yields an ion track radius of $R_{SAXS}\approx 4.1$ nm with a narrow Gaussian size distribution with a width of $\chi=0.2$ nm. The comparison of the cylinder model with the experimental SAXS data further allows us to estimate the density change to be less than 1\% relative to bulk GaSb ($\rho_e=40.77\times10^{10}$ cm$^{-2}$ taken from Irena \cite{irena}).\\
\indent
The track radius determined by SAXS matches the results from RBS measurements very well, as channeling is expected to be largely suppressed for the highly defective thin GaSb film on InP. No clear evidence of ion tracks was observed at lower fluences, most likely due to the very weak contrast between the amorphous ion tracks and the crystalline matrix. This explains the difficulties in studying ion tracks in GaSb by SAXS in our previous studies \cite{Kluth2014}.\vspace*{-0.6cm}
\subsection{Strain and ion hammering\vspace*{-0.3cm}}
\begin{figure}
  \includegraphics[width=\figsize]{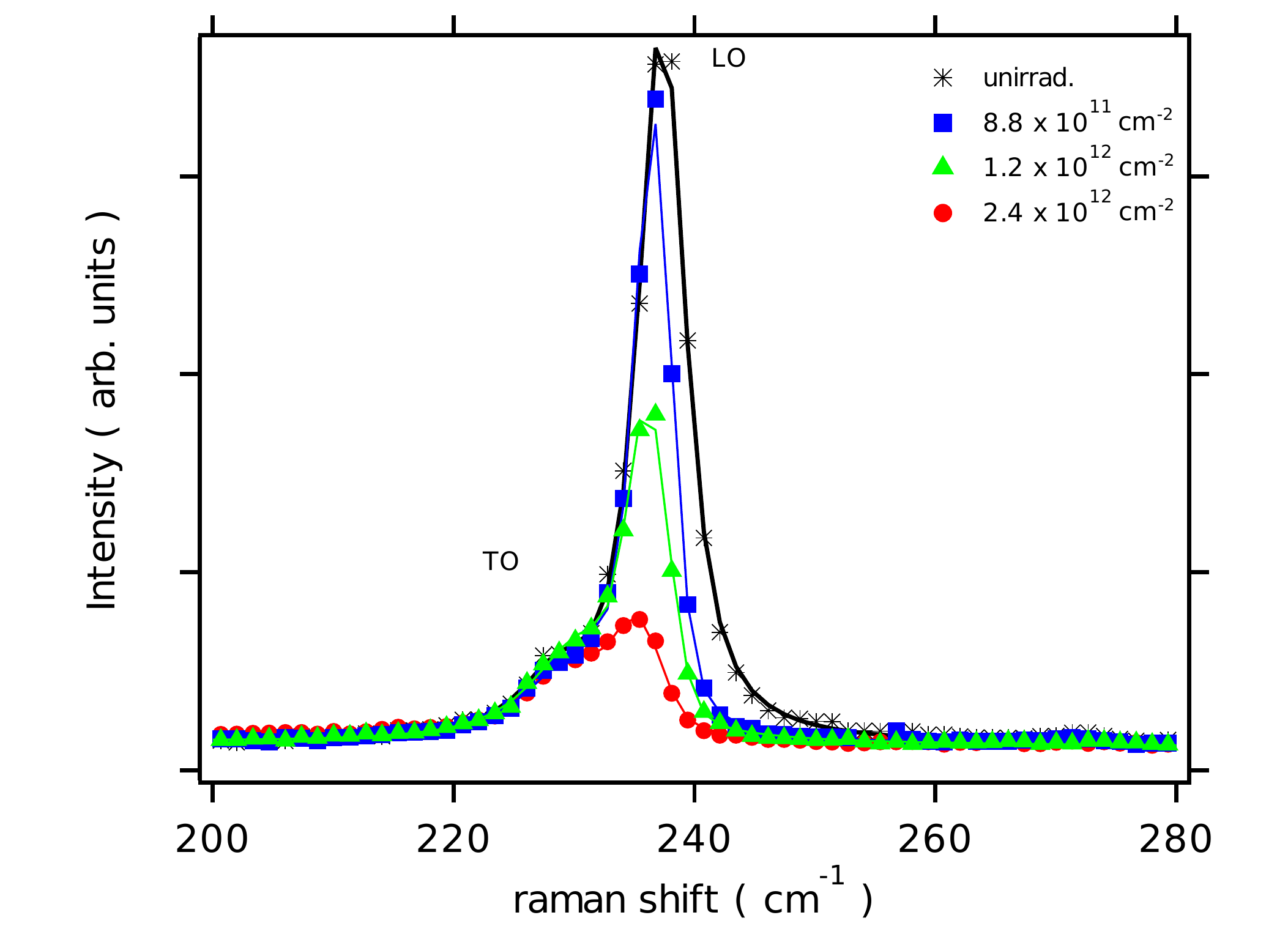}%
 \caption{Selected Raman spectra of GaSb samples irradiated at normal incidence with different fluences (symbols) and pseudo-Voigt fits to the data (solid lines).\vspace*{-0.5cm}\label{fig:raman_spec}}%
 \end{figure}
Figure \ref{fig:raman_spec} shows typical Raman spectra measured prior to the RBS/C experiments on the same samples (symbols) and fits to the data using pseudo-Voigt functions to describe the transversal optical (TO) and longitudinal optical (LO) Raman peaks (solid lines). The Raman spectra show an almost constant TO line intensity and a decreasing LO line, indicating increasing amorphisation with increasing fluence. The intensity reduction of the LO line is a measure of the amorphised volume fraction, and the fluence dependence is in good agreement with the RBS data.\\
\begin{figure}[b]
  \includegraphics[width=\figsize]{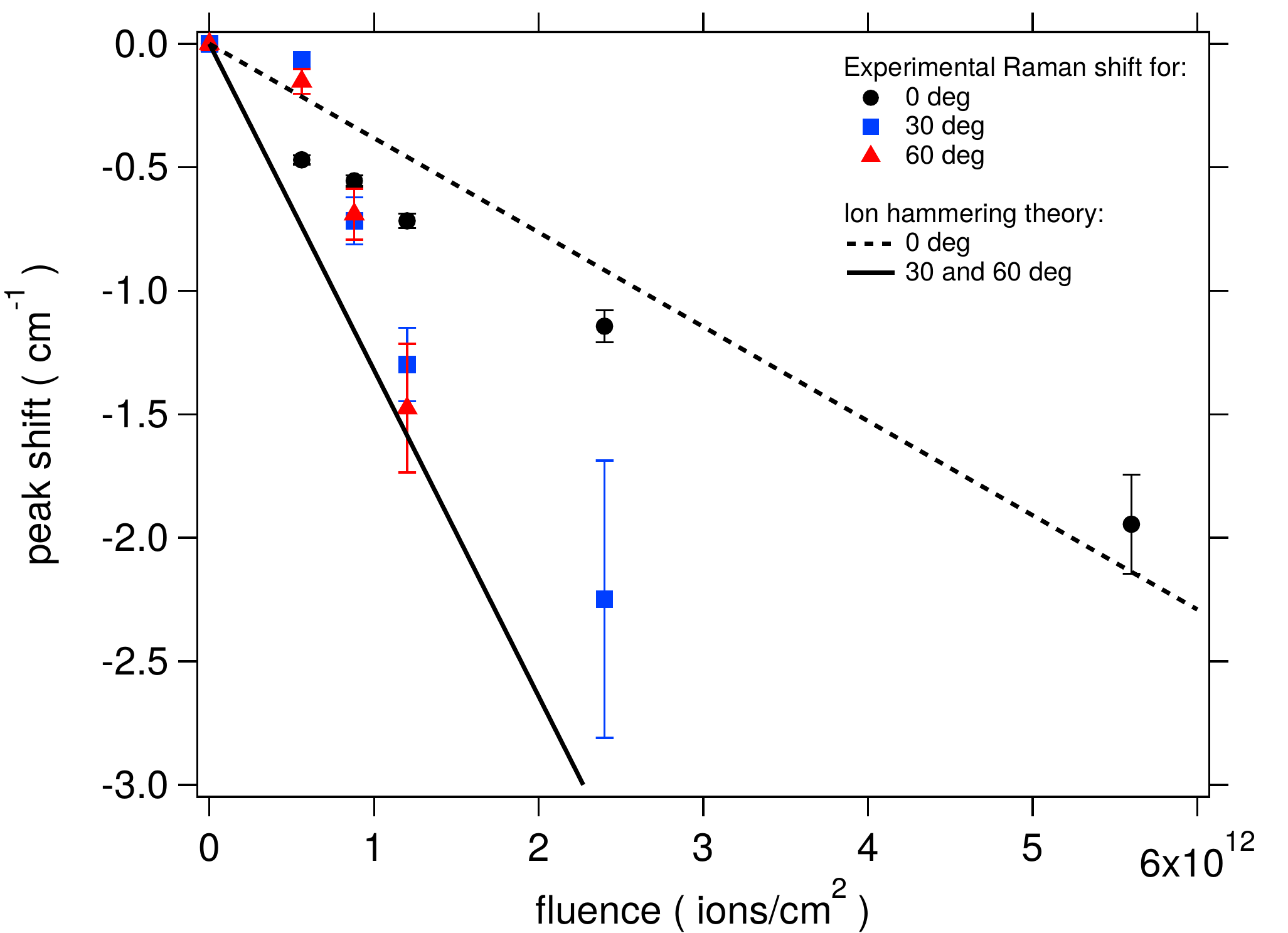}%
 \caption{Peak shift of the LO Raman line as a function of ion fluence for different irradiation angles relative to the surface normal (symbols) and peak shift predicted by ion hammering theory for irradiation normal to the surface (dashed line) as well as 30\degree~ and 60\degree~ (solid line).\vspace*{-0.3cm}\label{fig:raman_peak_shift}}%
 \end{figure}
\indent
In the fluence region of well-separated ion tracks in an otherwise crystalline matrix ($\Phi \ll 1\times10^{13}$ ions/cm$^2$), we observe a small but clear shift of the LO-phonon line towards smaller wave numbers, as shown in Fig. \ref{fig:raman_peak_shift}. This indicates the accumulation of tensile strain in the matrix. The strain increases almost linearly with fluence for all investigated irradiation angles. We observe the same slope for irradiation under 30$\degree$ and 60$\degree$ relative to the surface normal. Similar to the macroscopic surface shift observed in glasses and other semiconductors, we attribute the observed strain to the so-called ion hammering effect \cite{shift_apl_1997,shift_PRB_2005}; however, the crystalline matrix\clearpage \noindent responds with an elastic rather than a plastic deformation at low fluences $\Phi$.\\
\indent The ion hammering tensor is defined by \cite{PRL74,NIMPRB225,PRB72}
\begin{equation}
\small
\underline{\underline{\epsilon}}=A_0\Phi
\begin{pmatrix}
1-3 \sin^2(\theta) & 0 & 3 \sin(\theta)\cos(\theta)\\
0 & 1 & 0\\
3\sin(\theta)\cos(\theta) & 0 & 1-3\cos^2(\theta)
\end{pmatrix},\label{eq:ionhammering}
\end{equation}
where the deformation yield induced by a single ion track is given by
\begin{equation}
\small
A_0=1.164\frac{1+\nu}{5-4\nu}\frac{\alpha g S_e}{e \rho C} ,
\end{equation}
with the Poisson ratio $\nu=0.31$, density $\rho=5.61$ g/cm$^3$, specific heat $C=0.25$ J/gK, linear thermal expansion coefficient $\alpha=7.75\times10^{-6}$ K$^{-1}$ of GaSb at room temperature\cite{matpar}, and the fraction of energy transferred from the swift heavy ion to the thermal spike $gS_e$.\\
\indent
Within the framework of an analytical thermal spike model, the knowledge of the ion track radius allows us to estimate the efficiency of the energy transfer $gS_e$.\\ Following Szenes \textit{et al.} \cite{PRB65}, the ion track radius is defined by
\begin{equation}
\small
r_{spike}=a(0)\sqrt{\ln{\frac{gS_e}{\pi \rho C \Delta T a(0)^2}}},
\end{equation}
with the energy loss $S_e=22.3$ keV/nm, initial Gaussian width of the thermal spike $a(0)=11.2$ nm \cite{PRB65}, density $\rho=5.61$ g/cm$^3$, specific heat $C=0.25$ J/gK, and the difference between melting point and irradiation temperature $\Delta T=685$ K. Using the track diameter of 5 nm (3 nm) obtained by our RBS/C experiments, we can estimate the energy transfer to about 2.88 keV/nm (2.54 keV/nm), which corresponds to an efficiency $g$ between 0.11 and 0.13 .\\
\indent
The Raman peak shift can be calculated using the well-known secular equation whose solutions yield the frequencies of the optical phonons in the presence of strain \cite{PRB5}:
\begin{widetext}
\begin{equation}
\small
\begin{vmatrix}
p\epsilon_{xx}+q(\epsilon_{yy}+\epsilon_{zz})+\lambda & 2r\epsilon_{xy} & 2r\epsilon_{xz}\\
2r\epsilon_{xy} & p\epsilon_{yy}+q(\epsilon_{xx}+\epsilon_{zz})+\lambda & 2r\epsilon_{yz}\\
2r\epsilon_{xz} & 2r\epsilon_{yz} & p\epsilon_{zz}+q(\epsilon_{xx}+\epsilon_{yy})+\lambda
\end{vmatrix}=0~, 
\end{equation}
where $\lambda=\Omega^2-\omega_0^2$ and $\Omega \approx \omega_0+\lambda / 2 \omega_0$ with the unstrained phonon frequency $\omega_0$.\\
\indent Substituting the ion hammering tensor for $\epsilon_{ij}$ we obtain
\begin{equation}
\small
\Delta \Omega(\theta) = -\frac{\omega_0}{2}A_0\Phi(\frac{p-q}{2\omega_0^2}+\frac{3}{2\omega_0^2}\sqrt{(1-4\cos(\theta)^2+4\cos(\theta)^4)(p-q)^2+16\sin(\theta)^2\cos(\theta)^2r^2})\label{eq:ramanshift1}
\end{equation}
\end{widetext}
for the strain-induced phonon line shift. Evaluating Eq. (\ref{eq:ramanshift1}) for $\theta=0$\degree~ and converting to wave numbers yields
\begin{equation}
\small
\Delta \nu_{0} = -2\frac{(p-q)}{2\omega_0^2}\nu_0 A_0 \Phi~.
\end{equation}
In case of $\theta=30$\degree~and 60\degree, Eq. (\ref{eq:ramanshift1}) simplifies to
\begin{equation}
\small
\Delta \nu_{30,60} = -\left(\frac{1}{2}\frac{p-q}{2\omega_0^2}+\frac{3}{4}\sqrt{\left(\frac{p-q}{2\omega_0^2}\right)^2+3\left(\frac{r}{\omega_0^2}\right)^2} \right)\nu_0 A_0 \Phi~.
\end{equation} 
The solid and dashed lines in Fig. \ref{fig:raman_peak_shift} show the expected strain-induced peak shift for the different irradiation angles according to the ion hammering model using the parameters $(p-q)/2\omega_0^2=0.22$, $r/\omega_0^2=-1.08$, $\omega_0^2=1.84\times10^{27}$ sec$^{-2}$ from Ref. \onlinecite{PRB5}, and $gS_e=2.88$ keV/nm determined from the previously discussed RBS/C results. We note that there are no free parameters involved in the calculation, i.e., no parameters are adjustable to fit the experimental data.\\
\indent The slope of the observed Raman peak shift with fluence matches the prediction from the ion hammering model very well. In contrast to previous reports on the ion hammering effect in Ge and Si under swift heavy ion irradiation \cite{PRB83,nmat3,JPD42}, where the occurrence of a low-density liquid state is needed to explain the experimental observations, the ion hammering effect in crystalline GaSb is consistent with the simple thermal expansion model characteristic for glasses. This is also consistent with the low density contrast between the amorphous ion tracks and the crystalline matrix observed from the SAXS measurements.\vspace*{-0.2cm}
\subsection{Swelling and surface shift\vspace*{-0.3cm}}
Figure \ref{fig:sem_0deg_30deg_60deg}(a) shows cross-section SEM images of samples irradiated normal to the surface with fluences ranging from 1.2$\times10^{13}$ to 2$\times10^{14}$ ions/cm$^2$, reproduced from Ref. \onlinecite{Kluth2014}. At low fluences $\Phi\leq$ 1.2$\times$10$^{13}$ ions/cm$^2$ we observe the formation of separated, almost spherical voids. With increasing fluence (1.2$\times$10$^{13}$ ions/cm$^2<\Phi<$ 1.2$\times$10$^{14}$ ions/cm$^2$), the voids become elongated pockets and self assemble into columnar structures. At higher fluences the pockets become more irregular, and signs of fiberlike structures develop at the sample surface. The porousification of the material with increasing fluence is accompanied by a strong, mainly uniaxial swelling.\\
\indent
Figures \ref{fig:sem_0deg_30deg_60deg}(b) and \ref{fig:sem_0deg_30deg_60deg}(c) show cross-section SEM images of samples irradiated 30$\degree$ and 60$\degree$ relative to the surface normal with fluences ranging from 2.4$\times10^{13}$ to 1.2$\times10^{14}$ ions/cm$^2$. The first and most prominent difference compared to irradiation at normal incidence is the increased porosity and therefore the significantly larger swelling of the GaSb film, up to about twice that
{
\newpage
\onecolumngrid

\begin{figure}[b]
  \includegraphics[width=0.75\linewidth]{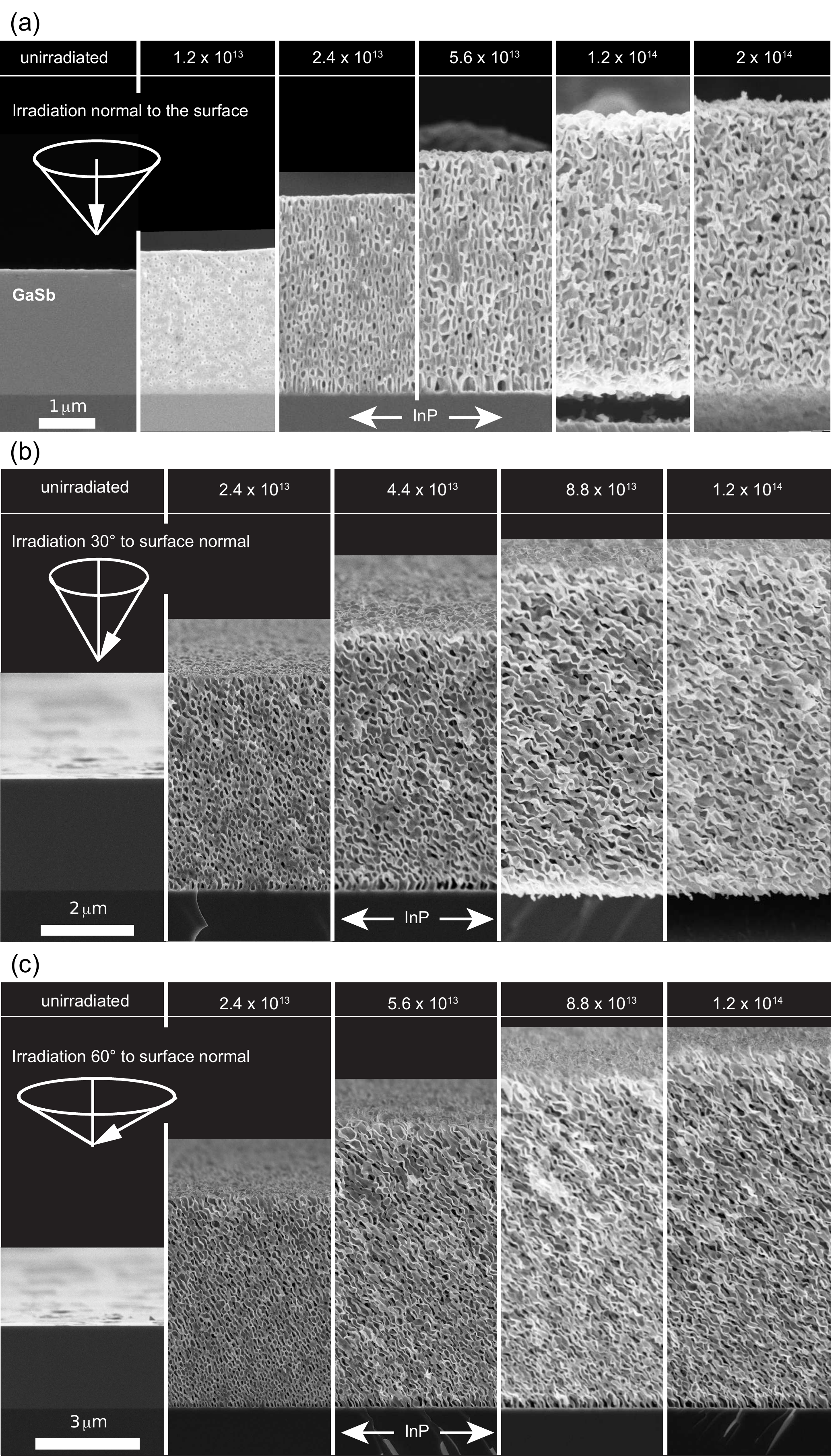}
 \caption{Cross-section SEM images of GaSb films on InP substrates irradiated normal to the surface (a), at an angle of 30\degree~(b), and 60\degree~(c) relative to the surface normal. The insets indicate the orientation of the ion beam during irradiation relative to the cross-sectional view.\label{fig:sem_0deg_30deg_60deg}}%
\end{figure}
\clearpage
\twocolumngrid
}
\newpage
\begin{figure}[h!]
  \includegraphics[width=\figsize]{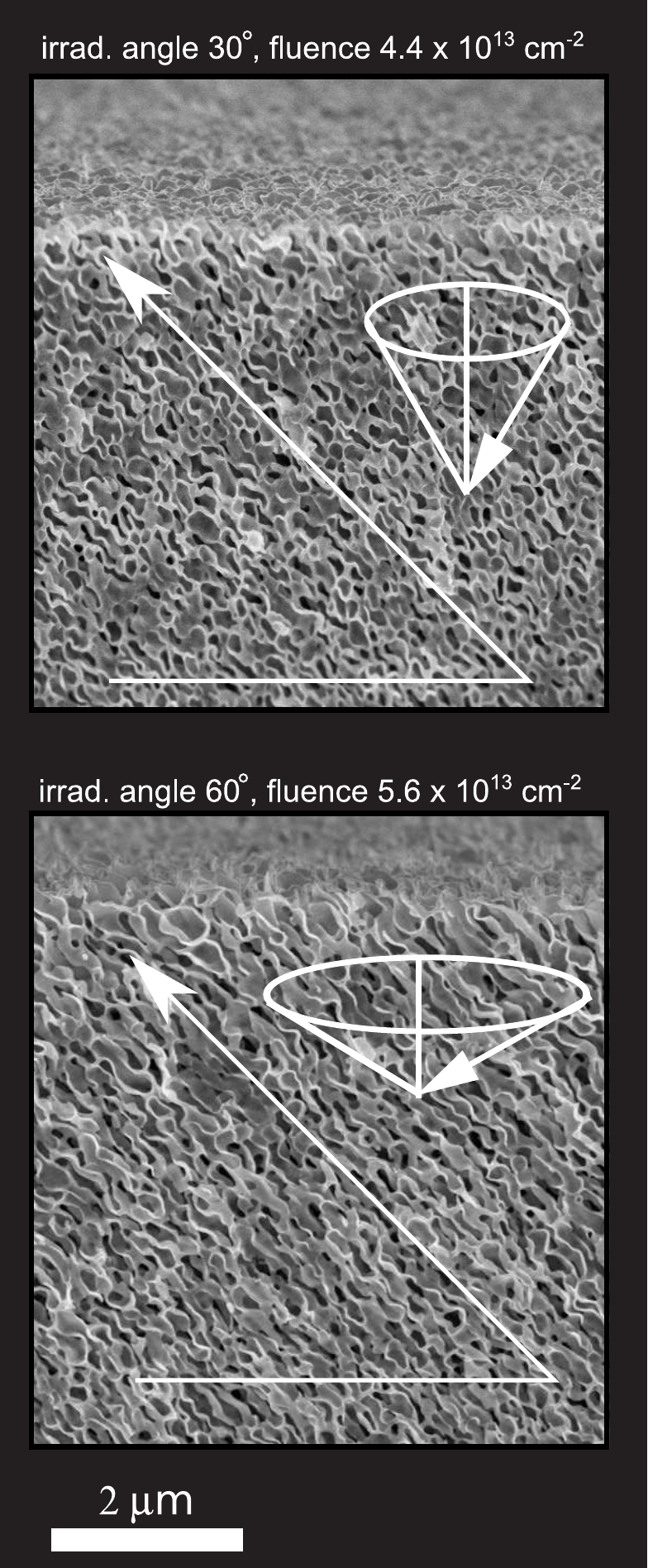}
 \caption{Cross-section SEM images of GaSb films on InP substrates irradiated with 185 MeV Au ions. Lines are a guide to the eye.\vspace*{-0.4cm}\label{fig:elongation}}%
\end{figure}%
\noindent
of normal incidence under similar irradiation conditions.\\
\indent
Another significant difference is a clear preferential orientation of the elongated pores as shown in Fig. \ref{fig:elongation}.\\
The pores are aligned about 45$\degree$ relative to the surface normal, and except for minor variations ($\pm5\degree$), the orientation is independent of the irradiation angle and fluence within the fluence range investigated. The elongation of pores with a preferential direction of about $45\degree$, essentially independent of irradiation angle and fluence, indicates the presence of considerable shear stress in the film \cite{neutron_irrad_voids}.\\
\begin{figure}
  \includegraphics[width=\figsize]{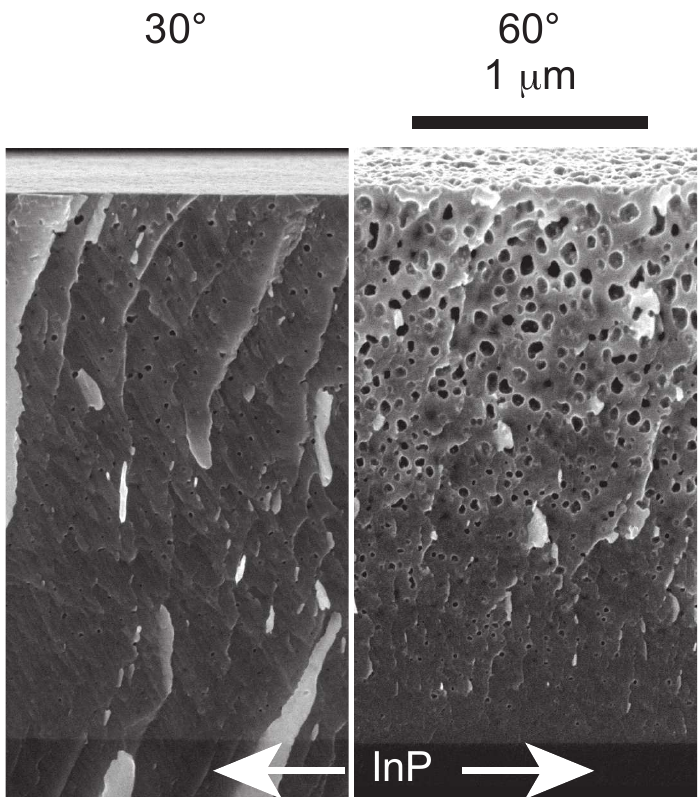}%
  \caption{Cross-section SEM images of GaSb films on InP substrate irradiated under 30\degree~and 60\degree~relative to the surface normal to fluence of $5.6\times10^{12}$ ions/cm$^2$.\vspace*{-0.5cm}\label{fig:sem_3}}%
\end{figure}
\begin{figure}
  \includegraphics[width=\figsize]{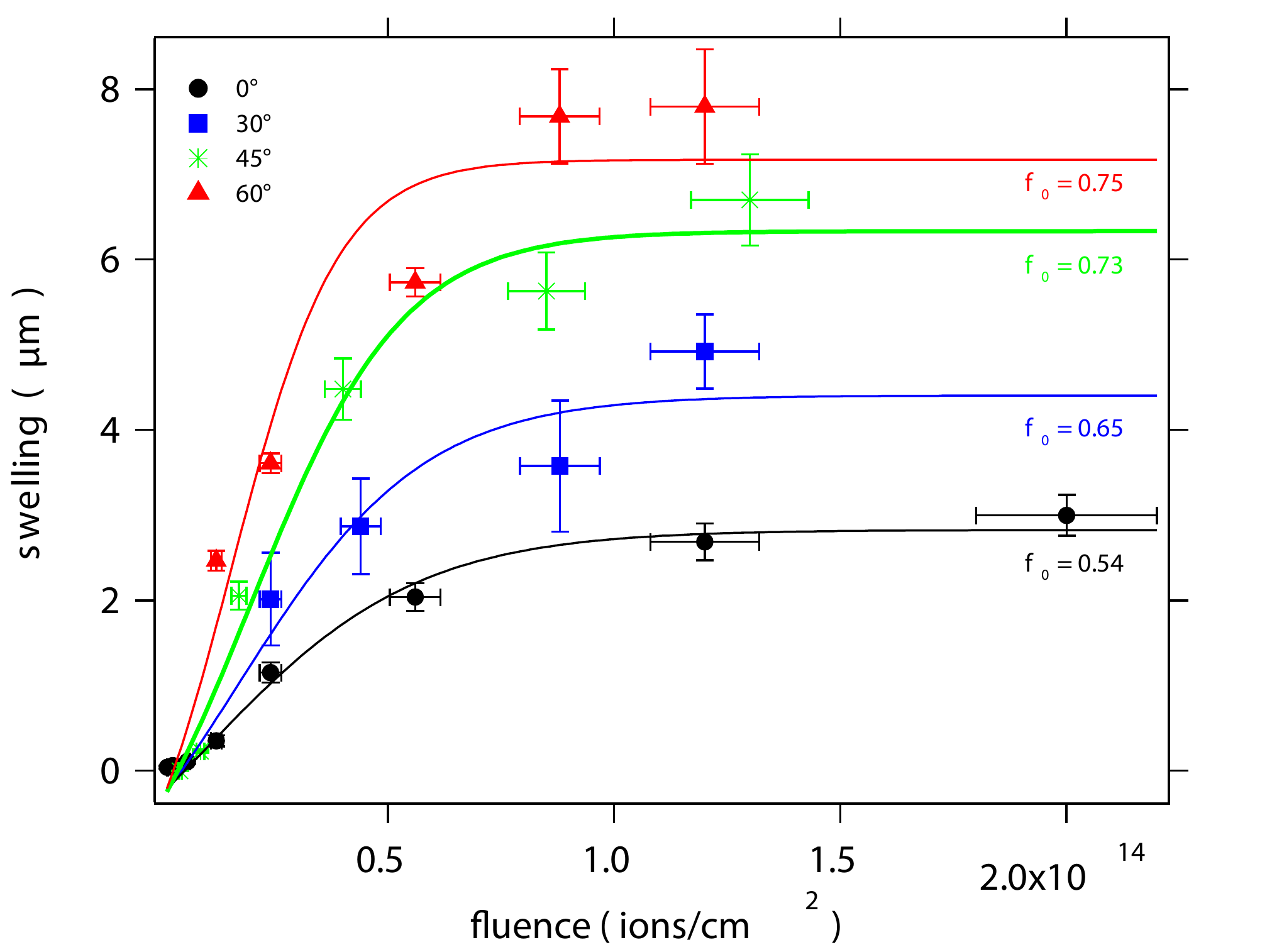}%
 \caption{Swelling of thin-film GaSb (determined from SEM cross-section images) as a function of ion fluence for different irradiation angles relative to the surface normal (symbols) and best fits of Eq. (\ref{eq:swelling2}) to the data (solid lines).\label{fig:swelling}}%
\end{figure}
Besides the preferential orientation, the high-fluence pore morphology is also significantly different to that at normal incidence. It changes from pocketlike structures to a structure resembling corrugated sheets, and the formation of fiberlike features at the surface can be observed at fluences as low as $\Phi=5.6\times10^{13}$ ions/cm$^2$. At low fluences, no discernible difference in pore shape can be observed, but a gradient in pore size becomes more apparent with increasing irradiation angle. SEM images in Fig. \ref{fig:sem_3} show a clear change from smaller pores at the interface to larger ones at the free top surface. Furthermore, it appears that the number density of pores also increases with increasing irradiation angle.
Figure \ref{fig:swelling} shows the swelling $\Delta h$ of the GaSb film as a function of fluence $\Phi$ for samples irradiated parallel (black dots), 30$\degree$ (blue squares) , 45$\degree$ (green stars), and 60$\degree$ (red triangles) relative to the surface normal. We have previously shown that a single ion impact into crystalline GaSb generates vacancy clusters which agglomerate/grow into larger voids by diffusion \cite{Kluth2014}. However, the formation of macroscopic voids and a measurable swelling is only observed above a threshold fluence $\Phi_0$ of about $5\times10^{12}$ ions/cm$^2$, which corresponds approximately to a full coverage of the sample with ion tracks and correspondingly to an almost complete amorphisation of the GaSb film, as observed by Raman and RBS/C.\\
\indent
The volume fraction of voids as a function of fluence $f_v(\Phi)$ can be described by a Johnson-Mehl-Avrami-Kolmogorov-type equation \cite{Avrami1,Avrami2,Avrami3}, assuming a nucleation and growth process similar to a crystallisation or phase-change process:
\begin{eqnarray}
\small
f_v(\Phi)=f_0{\bf[}1-e^{-\frac{\kappa}{ \cos(\theta)}(\Phi-\Phi_0\cos(\theta))}{\bf]}~,
\label{eq:swelling1}
\end{eqnarray}
with a correction for the ion path length due to the incident angle of the ions $\cos(\theta)$, the maximum/saturation porosity $f_0$, an effective ion cross section $\kappa$, and a threshold fluence $\Phi_0$. 
Assuming a uniaxial expansion only (base area $A=const.$), the porosity $f_v(\Phi)$ can be easily translated into the swelling $\Delta h(\Phi)$: 
\begin{eqnarray}
\small
\rho=\frac{m}{h A}=\rho_0(1-f_v(\Phi))~,\\
\Delta h(\Phi)=h(\Phi)-h_0=h_0\frac{f_v(\Phi)}{1-f_v(\Phi)}~,
\label{eq:swelling2}
\end{eqnarray}
with the bulk density $\rho_0$, the initial film thickness $h_0$, and an arbitrary but constant area $A$ and corresponding mass $m$.\\
\indent
The solid lines in Fig. \ref{fig:swelling} represent the best fits of Eq. (\ref{eq:swelling2}) to the data.
Based on RBS/C measurements presented before, the threshold fluence was chosen as $\Phi_0=5\times10^{12}$ ions/cm$^2$, which corresponds to between 80\% and 90\% amorphisation and corresponds approximately to complete coverage of the sample with ion tracks. The fits were performed in parallel, keeping the effective cross section $\kappa$ constant for all irradiation angles, which yields $\kappa=4.2\times10^{-14}$ cm$^2$. Only the saturation porosity $f_0$ is varied independently for the different angles as indicated in the graph. The significant increase in saturation porosity from $f_0=0.53\pm0.01$ for normal incidence to $f_0=0.75\pm0.01$ for 60\degree~irradiation angle is quite surprising. The intuitive explanation that the longer ion path results in greater vacancy production and clustering does not hold up, as this is implicitly included in Eq.  (\ref{eq:swelling2}) in the cosine term. In that case $f_0$ would be the same for all angles. Furthermore, if the total energy deposition in the layer was the only factor responsible for the swelling, a higher fluence for a lower angle should be able to compensate for the increased path length at higher angles, which is clearly not the case. At this stage we are not able to resolve the reason for the observed behaviour; however, possible mechanisms can include: (i) differences in macroscopic strain in samples irradiated under different angles, (ii) differences in mechanical stability of the different microstructures observed, and (iii) different vacancy production/nucleation rates. Possibly a combination of these effects may account for the observed behaviour.\\
As mentioned before, the strong elongation of the pores with a preferential direction of about 45\degree~as well as the observed Raman peak shift at low fluences indicate the presence of strong shear stress induced by the ion irradiation.\\
\begin{figure}
   \includegraphics[width=\figsize]{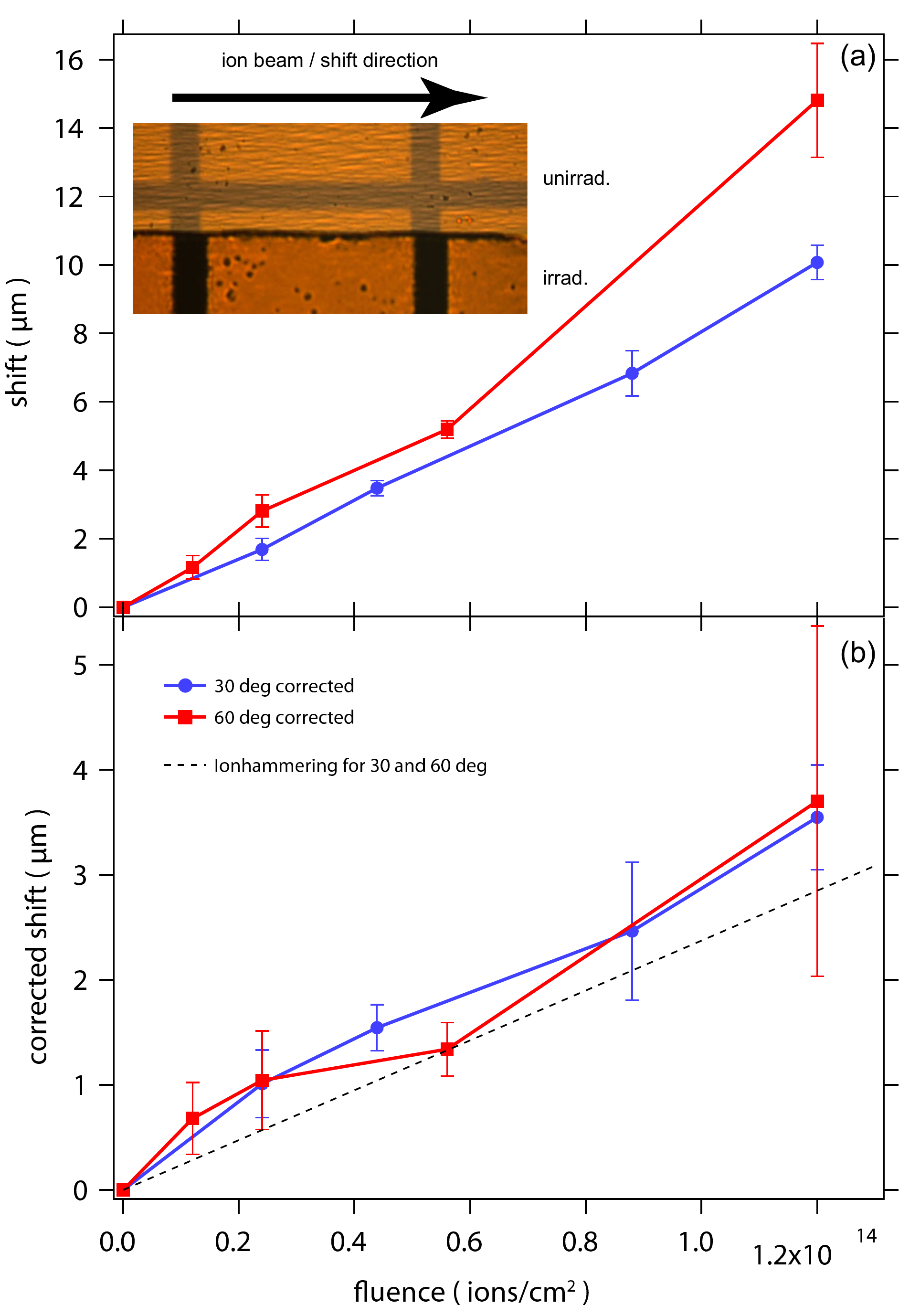}%
 \caption{Inset: Optical micrograph of a GaSb sample with Au markers irradiated under an angle of 60\degree~to a fluence of $5.6\times10^{13}$ ions/cm$^2$. (a) Surface shift as a function of fluence for 30\degree~(blue circles) and 60\degree~(red squares) irradiation relative to the surface normal. (b) Surface shift corrected for swelling as detailed in the text. \label{fig:shift}}%
\end{figure}%
\indent
We have prepared samples with 50 nm thick Au marker layers on the surface by thermal evaporation using TEM grids as a shadow mask and masked about half of the sample during swift heavy ion irradiation to investigate the plastic flow of GaSb under the induced shear stress. The inset in Fig. \ref{fig:shift} shows a typical optical micrograph of a sample with a Au marker layer where the lower part was irradiated under an angle of 60\degree~relative to the surface normal with 185 MeV Au ions to a fluence of $5.6\times10^{13}$ ions/cm$^2$, whereas the upper part was masked. The displacement of the Au squares in the irradiated area relative to the unirradiated top part clearly demonstrates a plastic flow in the direction of the ion beam projection on the sample surface (see arrow in the inset of Fig. \ref{fig:shift}). The observed surface shift $\Delta x$, determined using SEM images, is depicted in Fig. \ref{fig:shift} (a) as a function of fluence for 30\degree~(blue circles) and 60\degree~(red squares) irradiation relative to the surface normal. The strong swelling of the layers alters the observed surface shift and needs to be corrected for. Figure \ref{fig:shift} (b) shows the corrected surface shift after applying a simple geometric correction for the swelling by dividing the surface shift $\Delta x$ by the swelling $\Delta h$ using Eq. (\ref{eq:swelling2}). After correction, we observe the same deformation yield (slope) for samples irradiated at 30\degree~and 60\degree. Equation (\ref{eq:ionhammering}) allows for the estimation of the expected surface shift within the ion hammering model for irradiation under 30\degree~and 60\degree, shown as a dashed line in Fig. \ref{fig:shift}. Again, we note that there are no free parameters involved in the calculation. The good agreement between the experimentally observed surface shift and the ion hammering model indicates that even after full amorphisation and the development of significant porosity, the track formation process and the associated strain production resembles the results in the crystalline, low-fluence regime obtained by Raman earlier. In contrast to the low-fluence regime, where no strain relaxation is observed, it appears that the pre-damaged, amorphous layer relaxes the swift heavy-ion-induced strain completely through plastic flow similar to the ion hammering effect in glasses \cite{PRL74,BookWesch}.\\
\indent
This result is quite surprising as the melting point density of GaSb\cite{Wang2006} is with $\rho_{liquid}=6.058$ g/cm$^3$ about 8\% more dense then the amorphous phase, and the ion hammering model predicts a plastic deformation with negative deformation yield if a simple liquid-solid phase transition (first-order phase transition) is assumed\cite{BookWesch}. For this situation a surface shift in the other direction would be expected for GaSb. A similar situation is observed in amorphous Si and Ge and was attributed to the existence of a low density liquid phase in Si and Ge, which explains the observation of a positive deformation yield\cite{BookWesch,PRB83,nmat3,JPD42}. In contrast to Si and Ge, GaSb is a compound semiconductor and the liquid phase is already a low density phase, characterised by the presents of local order\cite{Medvedev2019,Gu2005}.  There is no indication that an even lower density phase exists. There is experimental evidence for a change in viscosity in the GaSb liquid phase; however, no evidence for a structural phase transition could be found (see e.g. Ref. \onlinecite{Gu2005}). The origin of the viscosity change is still under discussion. Our low fluence (separate amorphous tracks in a crystalline matrix) Raman results and the observation of a positive deformation yield at high fluence thus are both in good agreement with the ion hammering effect in glasses\cite{PRL74,BookWesch}.
\subsection{Swift heavy ion induced recrystallisation\vspace*{-0.3cm}}
\begin{figure}
  \includegraphics[width=\figsize]{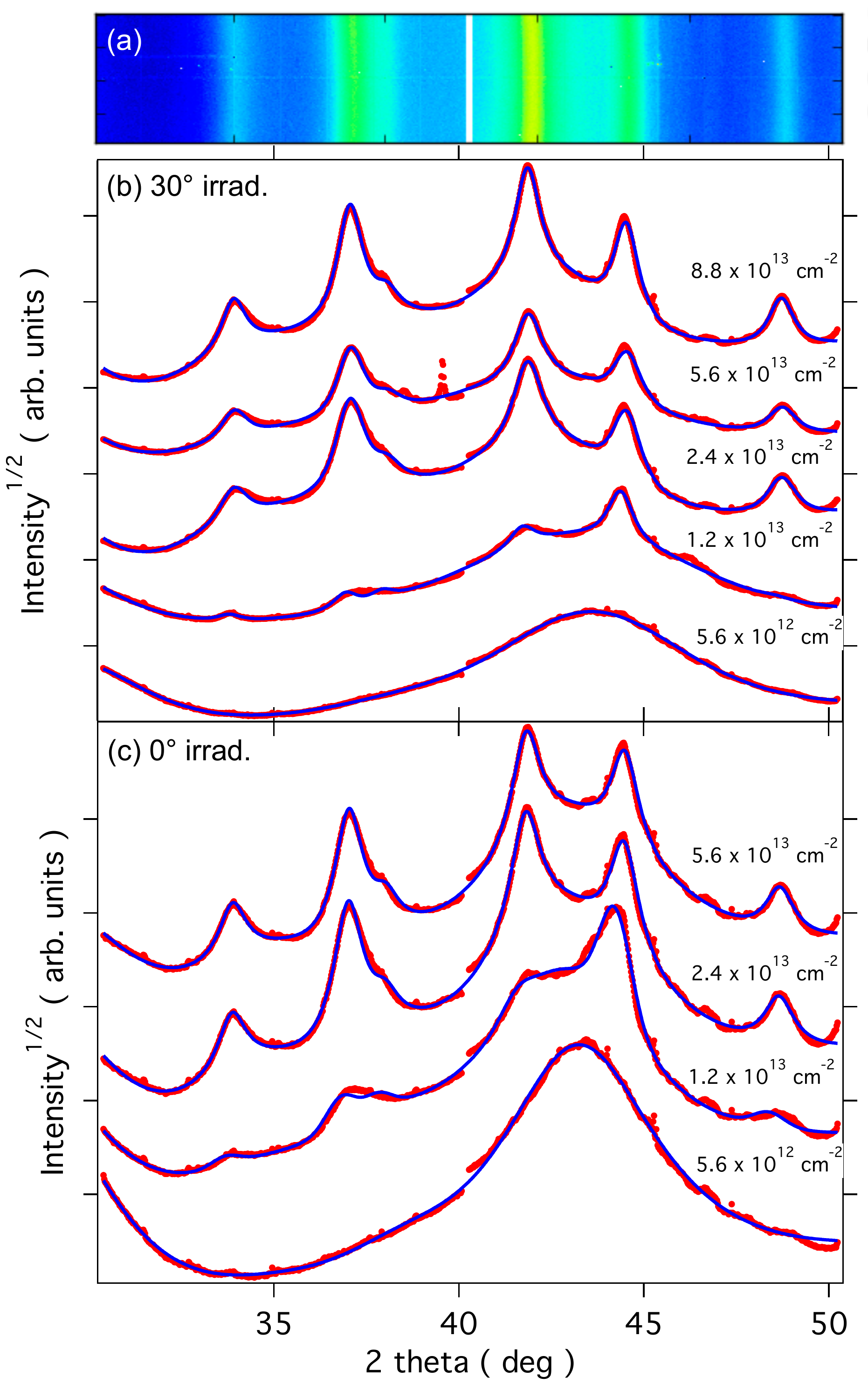}%
 \caption{Typical WAXS detector image (a), and azimuthally integrated scattering intensities (red dots) for samples irradiated with different fluences at 30$\degree$ relative to the surface normal (b) and at normal incidence (c). Solid (blue) lines are best fit Rietveld refinements to the experimental data.\label{fig:waxs}}%
\end{figure}
Figure \ref{fig:waxs} (a) shows a typical detector image from the WAXS measurements performed on thin-film GaSb samples in grazing incidence. For quantitative analysis, the images are azimuthally integrated and fitted by Rietveld refinement using a modified code based on MStruct \cite{mstruct,objcryst}, were we have implemented size and strain broadening following Refs. \onlinecite{JAC34} and \onlinecite{JAC26}, respectively. Figure \ref{fig:waxs} shows the integrated scattering intensities (symbols) and Rietveld fits (solid lines) for a sequence of samples irradiated with fluences from $5.6\times10^{12}$ to $5.6\times10^{13}$ ions/cm$^2$ at normal (c) and $30\degree$ (b) incidence ion irradiation relative to the surface normal. Consistent with our RBS/C experiments, both samples irradiated with a fluence of $5.6\times10^{12}$ ions/cm$^2$ show no sign of a crystalline phase; only a broad undefined background (most likely from the GaSb/InP interface and the InP substrate) can be observed. With increasing fluence, well-defined diffraction peaks become visible, which can be refined using a single nano-crystalline GaSb phase. (GaSb at ambient condition is a zinc-blende structure, which belongs to the space group F-43m in the Hermann–Mauguin notation, and atomic position (0,0,0) and (0.25,0.25,0.25) for Ga and Sb with occupancy 1, respectively) \cite{JAP81}.) The Rietveld refinement yields a lattice constant of $a=6.09$ \AA~ and $a=6.08$ \AA~ for irradiations normal to the surface and 30$\degree$ inclined, respectively. The lattice constant matches the 6.096 \AA reported in Ref. \onlinecite{JAP81}, surprisingly well considering the stress/strain induced by the ion hammering effect. This is a strong indication that the ion-induced stress is fully relaxed into the macroscopic, plastic deformation of the GaSb film. Within the data quality there is no indication for microstrain broadening typically observed in nanoparticles. Amorphous and crystalline GaSb have a very similar local structure (fourfold coordinated) and density \cite{Kalkan2013}. Therefore, it is most likely that the nano crystals embedded in an amorphous matrix with the same local structure do not have to minimise their surface energy through surface reconstruction or microstrain. The Rietveld refinement yields a crystallite size of 4.6 nm for irradiation normal to the surface and a slightly larger size of 5.3 nm when irradiated at 30$\degree$. Interestingly, in both cases the crystallite size is independent of the fluence, once the nanoparticles are visible in the diffraction pattern. Swift heavy-ion-induced annealing and recrystallisation has been observed previously in other materials (see, e.g., \onlinecite{Debelle2019,Hlatshwayo2016,Mihai2019,Som2005}); however, in GaSb we observe the formation of nanoparticles during a single ion impact, and no further crystallite growth is induced upon consecutive ion impacts. A likely explanation for the observation of nanoparticle formation only after the onset of porosification as well as the small but clear difference is nanoparticle size between 0\degree~and 30\degree~irradiation is the reduction in thermal conductivity due to the porous structure which causes higher temperatures on a longer time scale after a swift heavy ion impact and allows for the spontaneous nucleation of nanoparticles.
\section{Summary and conclusion\vspace*{-0.3cm}}
We have investigated the track formation in crystalline GaSb exposed to 185 MeV Au ions using Raman, RBS, and SAXS. The RBS measurements yield a track radius of about 5 nm for off-normal and 3 nm for normal-incidence ion irradiation, while about 4 nm is obtained from the SAXS data. We attribute the variation in track radius from RBS/C to a reduced energy loss due to channeling at normal incidence. The SAXS result for normal incidence is in between the RBS results for off-normal and normal irradiation and is most likely explained by suppressed channeling due to a high lattice mismatch of GaSb and InP and hence a high defect density compared to the single-crystal wafers used in RBS experiments. An alternative explanation for the difference in radius determined with SAXS and RBS/C lies in the fact that SAXS is sensitive only to density changes while RBS/C may also detect a damaged halo around the core track and overestimate the defect concentration due to the contribution of dechanneling, leading to systematically higher track radii measured by RBS/C. The different experimental methods consistently yield a larger track radius compared to the only other experimental report on track radii in crystalline GaSb by Szenes \textit{et al.} \cite{PRB65}. Raman spectroscopy was used for the investigation of ion-irradiation-induced strain in crystalline GaSb. Within the framework of ion hammering and assuming a simple thermal spike model, the known track radius was used to predict the ion-irradiation-induced stress in the crystalline GaSb sample. The prediction of the ion hammering/thermal spike model agrees very well with the stress observed by Raman spectroscopy.\\
\indent
Above a threshold fluence of about $5\times10^{12}$ ions/cm$^2$, which corresponds to an almost complete amorphisation of the GaSb film, the vacancy mobility and density are sufficient to allow macroscopic void nucleation and growth. This leads to significant porosity along with a strong swelling of the irradiated area. The degree of porosity and the resulting micro structure observed depend strongly on the irradiation angle. A significant increase in saturation porosity is observed for off normal irradiation. In addition to swelling, off-normal irradiation leads to significant plastic flow in the direction of the ion beam. The observed surface shift as a function of fluence for 30\degree~and 60\degree~irradiation matches well with the prediction based on the ion hammering model. We note that there are no free parameters involved in the ion hammering model to match the Raman and shift data. Wide-angle X-ray scattering shows the formation of nano-crystals once macroscopic porosity is observed, while the material is amorphised completely upon irradiation before porosity is induced. Rietveld refinement of the WAXS data reveals that the crystallite size is independent of the fluence but increases with increasing irradiation angle. This indicates that the nano crystals are formed during a single ion impact, rather than growing continuously with fluence.\\
\indent
The results reveal a complex transformation of the GaSb upon irradiation. First the material is rendered amorphous until completely covered by ion tracks. Subsequently, the material becomes porous where the microstructure and extent of the porosity are strongly dependent on the incident irradiation angle. After the formation of porosity, nano-crystallites form in the porous structure. Both at low and high fluences, results clearly show that in contrast to Ge and Si \cite{PRB83,nmat3,JPD42}, GaSb (crystalline and amorphous) behaves more like a glass where the ion-induced thermal spike locally melts the material and shear stress due to the thermal expansion in the molten track freeze-in during the subsequent rapid cooling. The formation of nano-crystallites can be explained by a spontaneous nucleation during the quenching of the ion-induced liquid phase inside the center of the thermal spike. Without porosity, the maximum temperature and lifetime of the thermal spike is too short to allow for crystallites to nucleate. With increasing porosity the thermal conductivity decreases locally, causing an increase in maximum temperature and lifetime of the thermal spike, which allows spontaneous nucleation and growth of nano-crystalline GaSb in the otherwise amorphous matrix.\\
\indent
The results are interesting from a fundamental point of view, as they show the glass like behaviour of the material, which clearly differs from that previously observed for elemental semiconductors. From an application point of view, the controlled formation of porosity offers the ability for the fabrication of thermoelectric and sensor devices.
\begin{acknowledgments}
We would like to thank the German Research Foundation (DFG) and the Australian Research Council (ARC) for financial support, and the staff at the ANU Heavy Ion Accelerator Facility for their continued technical assistance.  We also acknowledge access to facilities and technical support funded under the National Collaborative Research Infrastructure Strategy (NCRIS). These include the Heavy-Ion Accelerator Facility, the Australian Facility for Advanced Ion-Implantation Research (AFAiiR), and the ACT node of the Australian National Fabrication Facility.  Part of the research was undertaken on the SAXS/WAXS beamline at the Australian Synchrotron, part of ANSTO.
\end{acknowledgments}
\bibliography{mybib}
\end{document}